\documentclass[preprint,longbibliography,nofootinbib,12pt]{revtex4-1}
\usepackage{amsmath}
\usepackage{graphicx}
\usepackage{epsfig}
\usepackage{hyperref}
\usepackage{ulem}
\hypersetup{linktocpage,,colorlinks=true}
\setcitestyle{authoryear,aysep={:},citesept={;}}
\newlength{\upit}\upit=0.1truein

\newcommand{\ltappr}{{{\lower4pt\hbox{$<$} } \atop \widetilde{ \ \ \ }}}
\newlength{\bxwidth}\bxwidth=1.5 truein

\newlength{\figwidth}
\newlength{\shift}
\newcommand{\fg}[3]
{
\begin{figure}[ht]

\vspace*{-0cm}
\[
\includegraphics[width=\figwidth]{#1}
\]
\vskip -0.2cm
\caption{\label{#2}
\small#3
}
\end{figure}}

\newcommand \bea {\begin{eqnarray} }
\newcommand \eea {\end{eqnarray}}
\newcommand{\bk}{{\bf{k}}}

\begin{document}
\title{Emergence and Reductionism: an awkward Baconian alliance
}

\author{Piers Coleman$^{1,3}$}
\affiliation{
$^{1}$Center for Materials Theory, Department of Physics and Astronomy,
Rutgers University, 136 Frelinghuysen Rd., Piscataway, NJ 08854-8019, USA}
\affiliation{$^{2}$ Department of Physics, Royal Holloway, University
of London, Egham, Surrey TW20 0EX, UK.}
%\date{}
\begin{abstract}
This article discusses the relationship between 
emergence and reductionism from the perspective of a condensed matter
physicist.  
%, written as a contribution
%to the Durham Emergence Project
Reductionism and emergence play an intertwined role in the 
everyday life of the physicist, yet we rarely stop 
to contemplate
their relationship: indeed, the two are 
often regarded as 
conflicting world-views of
science. I argue that in practice, they compliment
one-another, forming an awkward 
alliance in a fashion envisioned
by the philosopher scientist, Francis Bacon.
Looking at the historical record in classical and quantum
physics, 
I discuss how emergence fits into a 
reductionist view of nature. 
Often, a deep understanding of reductionist
physics depends on the understanding of its emergent consequences. 
Thus the concept of
energy was unknown to Newton, Leibniz, Lagrange or Hamilton,
because they did not understand heat.  Similarly, the understanding of
the weak force awaited an understanding of the Meissner effect in
superconductivity. Emergence can thus be likened to an 
encrypted consequence of reductionism.
Taking examples from current
research, including topological insulators and 
strange metals, I show  that 
the convection between 
emergence and reductionism continues to provide a powerful driver
for frontier scientific research, linking the lab with the cosmos. 
\\

Article to be published by Routledge (Oxford and New York) as part of a volume entitled ``Handbook of
Philosophy of Emergence'', editors Sophie Gibb, Robin Hendry and Tom
Lancaster. Publication date June 2018. Copyright (\copyright Piers Coleman)
%We do great things.  
%$\langle \bar b \vert  = \langle 0 \vert  e^{\bar
%b\hat b \dg }$
\end{abstract}

%\eject
%
\maketitle
%
%\vfill\eject 

\section{Introduction: Reductionism and Emergence
}\label{}

Reductionism is the marvelous idea that as we take matter
apart to its smallest constituents, and understand the laws and forces
that govern them, we can understand 
everything.  This bold idea traces back to Greek
antiquity and has served as a key inspiration in the natural sciences,
particularly physics, 
up to the present day.  Emergence, by contrast, is the intriguing
idea that as matter comes together, it develops 
novel properties and unexpected patterns of collective behavior
\footnote{ For a careful discussion of the definition of emergence in physics,
see for example \cite[]{kivelson}.}.
This is something that scientists have long understood intuitively -
we observe emergence all around us - 
from snowflakes floating on a cold day, the pull of a mundane refrigerator
magnet, a flock of geese flying overhead, 
for those of us who have seen it, 
the magic of a levitating superconductor and life in all its
myriad forms. These are all
examples of natural science that 
that are not self-evident linear extrapolations of the microscopic
laws and which often require new concepts for their understanding.

Emergence and reduction are sometimes regarded as opposites.  The
reductionist believes that all of nature can be reduced
to a ``final theory'', a viewpoint expressed beautifully in
Stephen Weinberg's ``Dreams of a final theory''
\citep{Weinberg:1992nd}.  
Whereas reductionism is an ancient concept, 
the use of the word emergence in the physical sciences is a
comparatively recent phenomenon, dating back to the highly influential
article by Philip W. Anderson, entitled ``More is
Different'' \citep{emergence}. In this highly influential work, Anderson put
forward the idea that 
each level in our hierachy of understanding of science involves
emergent processes, and that moreover, the notion of ``fundamental''
physics is not tied to the level in the hierachy. 
Yet despite the contrast
between these two viewpoints, 
neither repudiates the other. 
Even the existence of a final
theory does not mean that one can go ahead and calculate its
consequences ``ab intio''. Moreover, the existence of emergence is not
a rejection of reductionism, and in no way implies a belief in
forms of emergence which can never be simulated or 
traced back to their microscopic origins.

In this article I present a pragmatic
middle-ground: arguing that reductionism
and emergence are mutually complimentary and quite possibly inseparable.
Sometimes methods and insights gained from
a  reductionist view, including computational simulation, 
do indeed enable us to understand collective emergent
behavior in higher-level systems.
However, quite frequently
an understanding of  emergent
behavior is  needed 
to gain deeper insights into reductionism.
The ultimate way to gain this deeper insight is through experiment,
which reveals the unexpected consequences 
of collective behavior amongst the microscopic degrees of freedom. 
In this way, the empirical
approach to science plays a central role in the intertwined
relationship between emergence and reductionism. 
This connection between reductionism, emergence and
empiricism lies at the heart of modern physics. 

Here, I will illustrate this viewpoint with examples from 
the historical record and also some current open challenges
in condensed matter physics. 

\section{The  Baconian View.}\label{}

Four-hundred years ago,  the  Renaissance philosopher-scientist, 
Francis Bacon championed a shift in science from the top-down
approach favored in classical times, to the empirically-driven
model that has been so successful up to the current day. 
In 1620, Bacon wrote\citep{Bacon:1994wu}
\begin{quote}
{\sl `` There are and can exist but two ways of investigating and
discovering truth. The one hurries on rapidly from the senses and
particulars to the most general axioms, and from them, as principles
and their supposed indisputable truth, derives and discovers the
intermediate axioms. And this way is now in fashion. The other derives
axioms from the senses and particulars, rising by a gradual and
unbroken ascent, so that it arrives at the most general axioms last of
all. This is the true way, but as yet untried. }\\
Francis Bacon, {\sl Novum Organum, Book 1, Aphorism {\bf XIX}, (1620)}
\end{quote}
Bacon argues for an 
integrated experimental-theoretical approach to science, in which
progress stems not from imposing the most general axioms, but by using
experiment and observation  without preconception, as guidance
to arrive at the ``most general axioms''. 
Bacon's approach
is not an abandonment of reductionism, but
a statement about how one should use experiment and observation to
arrive there.
The Baconian approach however leaves room for
surprises - for discoveries which are unexpected ``collective'' consequences of
the microscopic world, consequences which often shed new light on our
understanding of the microscopic laws of physics.  Bacon's
empirically driven approach plays a central role in the connection 
between emergence and reductionism. 

\subsection{The incompleteness of Classical Mechanics}\label{}

Modern education teaches classical mechanics as a purely reductionist
view of nature, yet historically it remained conceptually incomplete
until the nineteenth century, two hundred years after  Newton
and Leibniz.  Why? Because the concept of energy, a reductionist
consequence of classical mechanics, could not be 
developed until heat was identified as an
emergent consequence of random thermal motion.  This example helps us
to understand how emergence and reductionism are linked via experiment.

Energy is most certainly a reductionist consequence of classical
mechanics: if the force on a particle is given by
the gradient of a potential,
as it is for gravity, 
$\vec{F}= - \vec{\nabla} V = m
d\vec{v}/dt $, then from Newton's second law of motion, $\vec{F}= m
\frac{d\vec{ v}}{dt}$, one can deduce
the energy $E=
\frac{1}{2}mv^{2}+V $ is a constant of motion.  Moreover, this reductive
reasoning can be be extended to an arbitrary number of interacting
particles. 
Yet although Newton and
Leibniz understood the motion of the planets, understanding that was
considerably sharpened by Lagrange and Hamilton, the concept of energy was
unknown to them.  

Gottfried Leibniz had 
intuitively identified the quantity
$mv^{2}$ (without the half) as the {\sl life force} ({\sl ``vis viva''}) of a
moving object, but he did not know that it 
was the conserved counterpart of momentum ({\sl ``quantitas motus''}).
Lagrange\citep{Lagrange} introduced the modern notation
$T=\sum_{j}\frac{1}{2}m_{j}v_{j}^{2}$, with the factor of $1/2$,  and he
certainly knew that $T+V$ was conserved provided both are time-independent, a
point later formulated as a consequence of time-translation symmetry by
Emilie Noether in the 20th century\footnote{Lagrange
writes in his treatise {\sl M\'echanique Analitique} 
``In effect the integral $T+V=\hbox{constant}$ follows when $T$
and $V$ have no $t$ dependence'' ({\sl ``En effect, l'int\'egrale
$T+V=$const, ayant n\'ecessairement lieu, puisque 
$T$ \& $V$ sont fonctions sans $t$''})\citep{Lagrange} .}.
Yet still, the concept of energy had to wait a full two centuries after
Newton. 
From a practical point-of-view, momentum is a vector quantity which
is manifestly conserved in collisions, so that a macroscopic momentum 
can never dissipate into random microscopic motion.  
By contrast, energy as  a scalar quantity
inevitably transforms from manifest bulk kinetic energy, into microscopic
motion. 
Without an understanding of heat, kinetic energy appears to
vanish under the influence of friction. 

In Munich in 1798, 
the Colonial American-born royalist, inventor and physicist, Benjamin
Thompson (Count Rumford) carried out his famous experiment 
demonstrating that as a canon is bored, heat is produced.  He wrote afterwards\citep{Thompson1798} that
\begin{quote}
{\sl “It appears to me to be extremely difficult, if not quite impossible,
to form any distinct idea of any thing, capable of being excited and
communicated in the manner the Heat was excited and communicated in
these experiments, except be it MOTION.”}\\
Benjamin Count of Rumford, {\sl Phil. Trans. Roy. Soc. London {\bf
88}, p99 (1798).}
\end{quote}
Thompson's identification of heat as a form of motion eventually  put an end to
the ``caloric'' theory of heat as a fluid, clearing a
conceptual log-jam that had prevented progress for two centuries, 

From our 21st century hindsight, it seems almost inconceivable that
several generations of physicists would miss energy conservation.  Yet
nothing in science is ever simple. It was certainly known to Louis
Lagrange, and to William Rowan Hamilton after him, that 
the ``Hamiltonian''  $H=T+V$  is constant, provided that $T$ and $V$ have
no explicit time dependence, but the notion of the universally
conserved quantity we now call energy is completely absent from their
theoretical treatises. In his treatise of 1835\cite{hamilton1835}
in which William Hamilton 
introduces the concept of phase space and 
modern Hamiltonian dynamics, he explicitly comments that $H$ is
constant because $dH/dt=0$ (Equation 31 in \cite{hamilton1835})
but the significance of this constancy is not
discussed and Hamilton simply refers to it by its symbol, ``$H$''. 
In fact, though the word {\sl energy} was most probably first introduced by Thomas
Young in 1802\citep{Young1807}, the common usage of this concept had to wait until the
middle of the 19th century.

The modern reductionist might argue
that the early Newtonian physicists were just not reductionist {\sl
enough}!  Perhaps, had they been so, 
they would have realized that the conservation law known for
simple systems, would apply microscopically throughout 
macroscopic objects. Yet, historically,
until it was clear that heat was a form of random motion, this
connection was not made.  

Newton, Leibniz, Lagrange and
Hamilton were the greatest minds of their generation, they believed
fundamentally in the power of reductionism, yet they failed to make
the link. Would a modern reductionist, without modern hindsight have
fared any better? 
The fact is that the concept of
energy was hidden from the most brilliant minds of the era and
was not unlocked from its reductionist origins until physicists
had understood one of the most remarkable emergent consequences of
classical mechanics: heat.
Classical mechanics thus provides a beautiful illustration of
the intertwined relationship of reductionism and emergence.

\subsection{Darwin-Maxwell-Boltzmann}\label{}

Biologists trace the idea of {\sl emergence} back to Charles Darwin's {\sl Origin
of the Species}, and the use of the term in science began in
biology. Already, in the 19th century, scientists struggled with the relationship between emergence
and reductionism. 
In the origin of the species, Darwin writes\citep{Darwin59}
\begin{quote}
{\sl “whilst this planet has gone cycling on according to the fixed law of gravity, from so simple a beginning endless forms most beautiful and most wonderful have been, and are being evolved.”\\
Charles Darwin, Origin of the Species, p 490 (1859)}
\end{quote}
Here one glimpses in Darwin's writings, the idea that emergence and
reductionism are connected. 
Around the same time that Charles Darwin was writing his opus, a young 
James Clerk Maxwell was trying to work out how Newton's laws could 
give rise to Saturn's rings. To describe the rings, 
Maxwell constructed what was in essence, an
early model for his theory of atomic motion.  
In his prize essay on
the theory of Saturn's rings, Maxwell\citep{Maxwell59} wrote
\begin{quote}
{\sl 
We conclude, therefore, that the rings must consist of disconnected
particles; these may be either solid or liquid, but they must be
independent. The entire system of rings must therefore consist either
of a series of many concentric rings, each moving with its own
velocity, and having its own systems of waves, or else of a confused
multitude of revolving particles, not arranged in rings, and
continually coming into collision with each other. \\

 James Clerk
Maxwell, On the stability of Saturn's Rings. p67 (1859)}
\end{quote}
Maxwell understood 
that the
properties of Saturn's rings were a collective consequence of 
collisions  between its constituent particles. Later, when he moved
from Aberdeen to London, he used the astronomic inspiration from 
Saturn's rings as a model to develop his molecular theory of 
gases.  At a time where the concept of an atom was as 
controversial as modern string theory, his particulate model for 
Saturn's rings provided a valuable launching pad for 
his derivation of  the kinetic theory of molecular motion. 

Maxwell, and Boltzmann after him, realized the importance of the
Baconian approach to science - and in particular, that the collective
motion of particles required new statistical approaches, inspired by observation
and experiment.   Here's a quote from Boltzmann in the early 20th century\citep{Boltzmann1905}: 
\begin{quote}
{\sl “We must not aspire to derive nature from our concepts, but must adapt the latter to the former... Even the splitting of physics into theoretical and experimental is only a consequence of methods currently being used, and it will not remain so forever.”\\

Ludwig Boltzmann, Popul\" are Schriften, p77 (1905)}
\end{quote}
Boltzmann pioneered a reductionist explanation of
thermodynamics and the field of statistical mechanics, yet 
it is clear he was strong believer in the 
importance of an empirically-based approach.

\section{From the angstrom to the micron.}\label{}

The vast discoveries in physics during the twentieth century, from the
discovery of the structure of the atom, to relativity and
quantum-mechanics, the successful prediction of anti-matter from
relativistic quantum mechanics and the
discovery of gauge symmetries that lie behind the standard model of
particle physics, are a monumental tribute to the power of
reductionism\citep{pais}.  Today, the well-known extensions of this frontier lie 
in the puzzles of dark matter and dark energy, the observation of
gravity waves, the confirmation of the 
Higg's particle in the standard theory and string theory with its
prediction of 10$^{500}$ alternate multiverses\citep{weinberg2}.  The 
excitement of this frontier is widely shared with
society, for instance, in Stephen Weinberg's ``Dreams of a Final
Theory'', Brian Greene's ``Elegant Universe''\citep{Greene} and
Hawking's ``Brief History of Time''\citep{Hawking}. These 
expositions capture the beauty and romance of discovery while
giving rise to a popular, yet false impression that the frontier of science is
purely reductionist and that the frontiers lie at the extremes 
sub-quark scale, the Planck mass and the first moments of the Big Bang.

Yet, this is only one element of today's physics frontier:  we need
only to look just
below the limits of
the optical microscope and classical engineering, 
at scales of order a
micron to find remarkable emergent physics that we barely
begin to understand.
This is the view expounded by Philip W. Anderson (Fig. \ref{fig1})  in his highly
influential article ``More is different''\citep{emergence},
\begin{quote}
\sl `The behavior of large and complex aggregations of elementary
particles, it turns out, is not to be understood in terms of a simple
extrapolation of the properties of a few particles. Instead, at each
level of complexity entirely new properties appear, and the
understanding of the new behaviors requires research which I think is
as fundamental in its nature as any other.''\\
Philip W. Anderson in {``More is Different''}, p 393 (1972).
\end{quote}
Anderson's article, and his subsequent writings helped
to crystallize the idea of emergence in the physical sciences. The
concept of emergence re-invigorated the field of solid state physics,
prompting the field to redefine itself under the broader title
``Condensed Matter Physics''.  
\figwidth=\textwidth
\fg{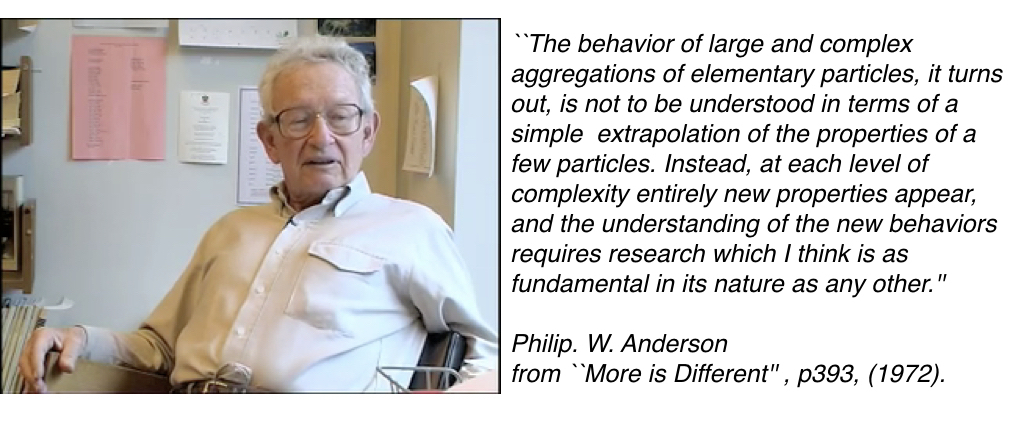}{fig1}{Philip W. Anderson. (Source:\ \href{http://musicofthequantum.rutgers.edu/video/ET_PhilipAnderson.mov}{musicofthequantum.rutgers.edu}) Anderson introduced the concept of
emergence into condensed matter physics  in his
influential ``More is Different'' article.  }

The terrestrial counter-part to the multiverse of string theory is the
periodic table. While there are only 92 stable elements, quantum
mechanics and chemistry mean that each new compound provides a new
universe of collective behavior.  As we go from elements to binary,
tertiary and quaternary compounds, out towards the organic molecules
of life, the number of unique combinations exponentiates
rapidly.  It is this emergent multiverse that provides the backdrop for quantum materials, biology,
life and all its consequences.

On the length-scale of atoms,
an Angstrom ($10^{-10}$m) we understand pretty much everything about
the motion of electrons and nucleii. This motion 
is described 
in terms of the Many-Body Schr\"odinger equation, which describes the
system in terms of a wave, described by the many-body wavefunction $\Psi
(1,2,3\dots  N)$, where $1, 2, \dots  $ denote the co-ordinates of the
particles.  
The squared magnitude of this wave provides the
probability of finding the particles at their respective co-ordinates,
\begin{equation}\label{}
p (1,2\dots N) = |\Psi (1,2\dots  N)|^{2},
\end{equation}
and in principle, with a few caveats, 
all the statistics of the particle motion, momentum,
energy, the fluctuations, correlations and response
can be determined from $\Psi $. 
One important aspect of this description, is its wave
character, reflected by the phase of the wavefunction.  When we add particles together, their waveforms
overlap and interfere with each other, so that unlike classical
systems, the probability distribution of the sum is not the 
sum of its parts
\begin{equation}\label{}
|\Psi_{A} +\Psi_{B} |^{2}\neq |\psi_{A}|^{2}+ |\psi_{B}|^{2}.
\end{equation}
This is part of the answer to something 
chemists know intuitively: that when one combines elements together,
the compound that forms is utterly different from a simple mixture
of its components. 
The other important aspect is that the wavefunction depends
on a macroscopically huge number of variables - classically, a system
of $N$ particles requires $3N$ position and momentum variables - a
quantity that is itself huge; yet 
quantum-mechanically, the number of variables required to describe 
a wavefunction is an exponential of this huge number. 

As we scale up from  the
Angstrom to the micrometer (1$\AA$ = 10$^{-10}$m, 1$\mu$m = 10$^{-6}$m ), a mere four orders of magnitude, 
matter acquires qualitatively new properties. 
The particles come together together to form crystals: this we can
understand classically. However, the electron waves that 
move throughout these immense periodic
structures interfere with each other and this interference
endows
 matter with remarkable new properties, hardness, rigidity,
magnetism, metalicity, semi- and superconductivity, phase transitions, topology
and much much more.  
To take an example proposed by Anderson\citep{emergence}, 
on the scale of the nanometer, the motion
of electrons in metallic gold is identical to that in niobium or tin.
Yet on scales of a micron, electrons in niobium
and tin correlate together into Cooper pairs  to form superconductors
that expel magnetic fields and levitate magnets.
Niobium and tin are examples of low-temperature superconductors,
requiring the extreme low temperatures of 
liquid helium to cool them to the temperatures where they
conduct without resistance, 
but today physicists have discovered new families of ``high
temperature superconductors'' that only require liquid nitrogen, and
there is a dream that room temperature superconductivity might occur in
hitherto undiscovered compounds. 
Yet superconductivity is just a beginning, for already by the micron, 
life develops.  The organism
Mycoplasma Mycoides, found in the human gut, forms self-reproducing
cells of 250nm in diameter\citep{kuriyan}. While we more-or-less understand the
physics of Cooper pairs in periodic, equilibrium superconductors, 
we are far from 
understanding the emergent physics of life that develops on the
same scale in aperiodic, non-equilibrium structures.  This lack of
understanding occurs despite our knowledge of the microscopic, many-body
Schr\"odinger equation, and it is this realization that prompts us to
appreciate emergence as a complimentary frontier\citep{middleway}.  It prompts us to
pose the question:
\begin{quote}
{\sl What are the principles that govern the emergence of collective behavior in matter?}
\end{quote}

\fg{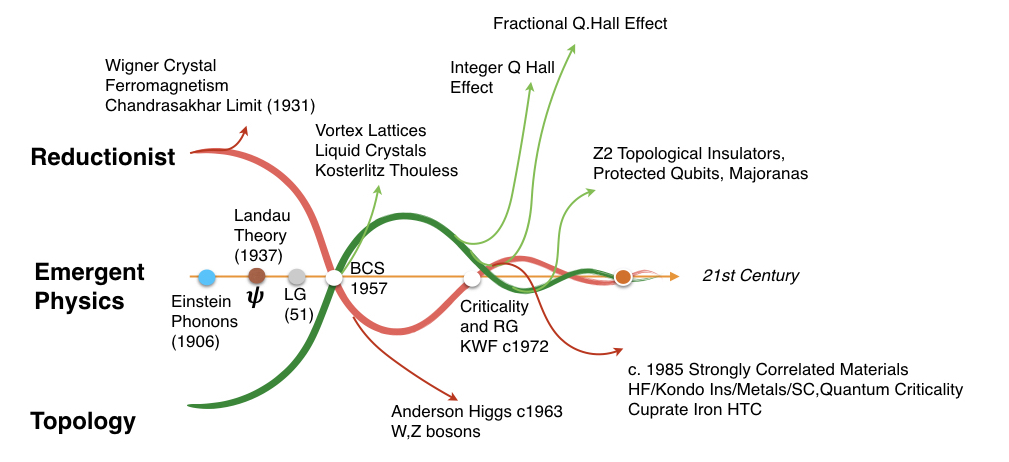}{fig2}{Schematic time-line,  illustrating
developments in condensed matter physics over the past century. The
three arrows show developments following a reductionist, emergent and
topological track.}

\section{A selective history of Emergence and reductionism in
Condensed Matter Physics}\label{}

Condensed matter physics is rife with historical examples of 
intertwined reductionism and emergence, with the one providing
insights into the other (Fig. \ref{fig2}). One of the things we learn from these
examples, is that fundamental physics principles are 
not tied to scale: that while
insights from the cosmos 
influence our understanding in the lab, equally, 
understanding of emergent 
principles gleaned from small-scale 
physics in the lab has given us extraordinary new insights into the
early universe and the sub-nuclear world. 

To illustrate this interplay between
emergence and reductionism,
let us look at some examples.
Quantum condensed matter physics arguably began with
Albert Einstein's 1906 proposal\citep{einstein2} that the concept of quanta could be extended from
light, to sound.  In the previous year he had proposed the idea of quanta, or
{\sl photons} to interpret Planck's theory of 
black-body radiation\citep{einstein1}. 
By proposing that light is composed of streams of indivisible
quanta of energy $E=hf$, where $h$=6.626$\times$10$^{-34}$Js is
Planck's constant and $f$ is the frequency, Einstein was able to
inject new physical insight into Planck's earlier work, 
and using it, he could make the link between black-body radiation and the
photo-electric effect.  By 1906 he saw that he could take 
the idea one step further,
proposing that analogous sound quanta occur in crystals. By treating a
crystal as an ``acoustic black-body'', Einstein was able to develop of
theory of the low temperature specific heat capacity of diamond, as it
drops below the constant value (``Dulong and Petit's law'') predicted
by classical equipartition.  Einstein's work in 1905 and 1906 are
remarkable examples of high-grade phenomenology - driven by experiment
and careful physical reasoning.  
%It took a further 20 years before Heisenberg was able to
%provide the appropriate a microscopic description 
%of the quantum harmonic oscillator, explicitly demonstrating energy quantization 
%and a full 40 years
%before a fully functioning quantum theory of electromagnetism
%became available.  
Moreover, Einstein's ``phonons'' as
we now call them, are emergent quanta of the solid state: the
result of the quantization of the collective motion of a macroscopic
crystal.

Another other early idea of emergence in physics, is
Landau's order parameter 
theory of phase transitions, developed in 1937\citep{landaupt}. Here he
introduced the key concepts of an order parameter and spontaneously
broken symmetry: the main idea is that the development of order 
at a phase transition
can be quantified in terms of {\sl order parameter}
$\psi $, which describes the development of a macroscopic
property, such as a  magnetization ($\psi = M$) or an electric polarization
($\psi =P$).  With a  very simple phenomenological
theory, Landau showed how to use this concept to describe phase transitions, 
without reference to the microscopic origin of the order parameter or
the mechanism by which it developed. In Landau's theory, close to a
second order phase transition,
the dependence of the bulk free energy $F[\psi ]$
on the order parameter is given by 
\begin{equation}\label{}
F[\psi ] = a (T-T_{c})\psi^{2}+ b \psi^{4} + O (\psi^{6})
\end{equation}
where $a$ and $b$ are positive constants, $T$ is the temperature and
$T_{c}$ is the critical temperature.  For $T>T_{c}$, the free energy
is a minimum at $\psi = 0$, but for $T<T_{c}$, it develops two 
``broken symmetry'' minima at $\psi = \pm [(a/2b) (T_{c}-T)]^{1/2}$
\citep{Chaikin95,Coleman16}.  The
important point about Landau theory, is that it describes a universal
property of matter near a phase transition, independently
of the microscopic details of the material.  Thirteen years later in
1950,  Ginzburg and Landau\citep{ginzburglandau} showed how an
more detailed version of Landau theory, or ``Ginzburg Landau theory'',
in which $\psi (x) $ is 
a complex order parameter with spatial dependence, could provide a
rather complete macroscopic description of superconductors accounting
for the expulsion of magnetic flux and the levitation of magnets a
half decade before the Bardeen Cooper Schrieffer (BCS) microscopic
theory of the same phenomenon. 

Yet condensed matter physics could not have developed without reductionism\citep{pais}.
With the arrival of Heisenberg's matrix mechanics in the 1920's, it
became possible to attempt a first-principles description of quantum
matter.  Suddenly, phenomena such as ferromagnetism that were
literally impossible from a classical perspective, could be given a
precise microscopic description, and these phenomena could be linked
in a reductionist fashion to the equations of quantum mechanics.  The
idea that electrons are probability waves, described by Schr\"odinger's
equation, led to the notion of Bloch waves: electron waves
inside crystals.  The idea of antimatter, predicted by Paul Dirac using his
relativistic theory of electrons\citep{dirac31} had its direct
parallel in condensed matter physics in 
Peierls' and Heisenberg's concept of
``hole'' excitations in semiconductors\citep{Hoddeson87,Heisenberg31}.  Landau and Ne\` el
extended Heisenberg's ideas of magnetism to predict
antiferromagnetism, first observed in the 1950s while Wigner used
reductionist principles to predict that electrons would form ``Wigner
crystals'' at low densities, a remarkable result not confirmed until
the 1980s.   Quantum mechanics also enjoyed application in the new
realm of astrophysics, most dramatically
in Subrahmanyan Chandrasakhar's
theory of stellar collapse\citep{RevModPhys.56.137}.  By combining classical gravity with the
statistical (quantum) mechanics of a degenerate fluid of protons and neutrons,
Chandrasakhar was able to predict that beyond a critical mass, stars
would become unstable and collapse. The critical Chandrasakhar mass
$M$ of a star,
\begin{equation}\label{}
M \approx M_{P}\left(\frac{M_{P}}{m_{p}} \right)^{2}
\end{equation}
is given in terms of the proton  
and the Planck mass, $m_{p}$  and 
$M_{P}=\left(\frac{hc}{G} \right)^{\frac{1}{2}}$ respectively.
Chandrasakhar's formula, built on principles designed to understand
the terrestrial statistical mechanics of electrons,  is the first time 
that gravity and quantum mechanics come together in a single expression.

Yet the fully reductionist revolution of quantum mechanics ran out of steam
when it came to understanding superconductivity: the phenomenon
whereby metals conduct electricity without resistance at low
temperatures.  Some of the greatest
minds of the first half of the 20th century, Bohr, Einstein\citep{sauer}, Bloch,
Heisenberg and Feynman\citep{schmalian} attempted microscopic theories of
superconductivity, without success. 
In 1957, the reductionist and emergent strands of condensed matter
physics, came together in a perfect storm of discovery, with the 
development of the Bardeen Cooper Schrieffer (BCS)
theory of superconductivity\citep{bcstheory}. 
On the one hand, it required
a reductionist knowledge of band theory and the interaction of
electrons and phonons; it also took advantage of the new methods of
quantum field theory, adapted from the theory of quantum electrodynamics by
early pioneers such as Fr\" ohlich, Gell-Mann and Hubbard. 
On the experimental front, it required the
discovery of the Meissner effect: the expulsion of magnetic fields
that occurs when a metal becomes superconducting; it also built strongly
on the phenomenological ideas of London, Landau and Ginzburg, Pippard and
Bardeen; finally,  it required
stripping the physics down to its bare minimum, in the form of a 
minimalist model now known as the ``BCS model''. The important point
is that rather than
attempting a fully reductionist 
description of the combined electron-lattice and electron-electron
interactions, which led to something far too complicated to be solved
in one go, Bardeen, Cooper and Schrieffer captured the the combined effects of these
phenomena in terms of a simple low-energy attractive interaction
between pairs. 

BCS theory had many further ramifications: pairing was 
generalized to the nucleus, where it led to an understanding of the
stability of even-numbered nucleii; it led to the
prediction of superfluidity in neutron stars and He-3. Most
unexpectedly, it opened up new perspective on broken symmetry that 
inspired Anderson then Higgs
and others to identify a mechanism
for how gauge particles acquire mass that we now call the
``Anderson-Higgs mechanism'' \citep{andersonhiggs1,peterhiggs64}.  At a time where particle
physicists had almost abandoned field theory, the new success in
superconductivity provided a case study of field theory in action 
that stimulated a resurgence of interest in
field theory in particle physics, 
leading to Electro-weak theory\citep{Witten16}.  Indeed, key elements of
electro-weak theory can be understood as a simple two component
spinorial extension of Landau Ginzburg theory, and from this
perspective, the weak force in nuclear-particle  physics 
can be understood as a
kind of cosmic Meissner effect that expels the W and Z fields from our
universe.

A second example of the intertwined nature of reductionism and
emergence is provided by the theory of critical phenomenon, a 
revolution in understanding of phase transitions that occurred 
a decade after BCS theory, between 1965-1975\citep{domb}.  From the
sixties, physicists were increasingly aware 
of a failure in the classical theory of phase transitions, 
based on the work of 
Van der Waals, Landau and others, which was
unable to described the observed properties of second order phase transitions. 
Experiments, plus 
and Onsager's solution to the two dimensional Ising model,  showed that
phase transitions were characterized by  
unusual, indeed, universal 
power-law behavior. For example, the magnetization of a
ferromagnet below its critical temperature develops with a power-law
$M\propto (T_{c}-T)^{\beta}$. 
Landau's theory predicts $\beta=1/2$,
yet in three dimensional Ising  ferromagnets, 
$\gamma = 0.326\dots $.  Moreover, the unusual critical exponents were
found to occur in a wide variety of different phase transitions,
exhibiting the phenomenon of ``universality''. 

To understand this discrepancy required a
revolution in statistical mechanics, involving new, high precision
measurements of phase transition, it meant borrowing methods that had been
developed to control or ``renormalize'' divergences in particle
physics, but it also involved
developing new ideas about how physics
changes and scales with size. Today these ideas are captured 
by a ``scaling equation'' that describes the evolution of a
Hamiltonian $H$ with length-scale $L$. Schematically, such scaling
equations are written as 
\begin{equation}\label{}
\frac{\partial H}{\partial Log [L]} = \beta [H], 
\end{equation}
where $H$ is the Hamiltonian, $L$ represents some kind of minimum cut-off
length-scale to which the Hamiltonian applies and $\beta [H]$, the
function that describes how $H[L]$ depends on length scale is called
the ``beta function''. 
The culmination of this work 
in Fisher and Wilson's ``epsilon expansion'', showed how to calculate 
scaling behavior using a beautiful
innovation of following physics as a function of dimension $d$\citep{PhysRevLett.28.240}. 
Remarkably, for the simplest models, 
the classical theories of phase transitions worked in dimensions above
$d=4$. Wilson and Fisher showed that a controlled expansion of the
critical properties could be developed 
in terms of the deviation from four dimensions
$\epsilon=4-d$.  The Fisher Wilson theory is a theory of an emergent
phenomenon, yet it draws on methodologies from 
reductionist quantum field theory. 

\section{Two examples from current physics}\label{}

The 
convective exchange of reductionist and emergent perspectives
continues to drive current developments in condensed matter physics. 
I'd
like to touch on
two active examples: research into 
{\sl topological} properties of quantum matter, and the
mystery posed by the discovery of 
classes of  phase transitions at absolute zero, which radically
transform the electrical properties of conductors into 
{\sl strange metals}.

\subsection{A topological connection}\label{}

One of the most remarkable developments
has been the discovery of a topological connection to emergence\citep{hasan_rmp2010,Moore:2010ig}.
Topology describes global properties of geometric
manifolds that are unchanged by continuous deformations.
For instance, a 
donut can be continuously deformed into a one-handled mug: the presence
of the hole, or the handle is topologically protected and we say 
they have the same topology. 
Mathematics links the
differential geometry of two dimensional manifolds to the topology
\[
\hbox{differential geometry}\leftrightarrow \hbox{topology}
\]
via the ``Gauss Bonet'' theorem, 
\[
\frac{1}{4\pi}\int \kappa dA = (1-g)
\]
which relates the area integral of the curvature to the number of
handles or the {\sl genus} $g$ of
the surface. Topology is a kind of mathematical emergence: a 
robust property that depends on the global properties of a manifold.

The rise of topology in condensed matter physics involved a
marvelously tortuous path of discovery. While the microscopic physics
is a reductionist consequence of the 
band-theory of  insulators developed in
the 1930s,  the discovery of a topological connection had to await another
half-century,  culminated in the discovery of a new class of band
insulator, the ``topological insulator''. One of the remarkable
properties of topological matter is that the surface remains
metallic.  The 2016 Nobel prize in physics to Haldane,
Kosterlitz and Thouless was awarded for their early contributions to
topology in condensed matter physics.

\figwidth=0.9\textwidth
\fg{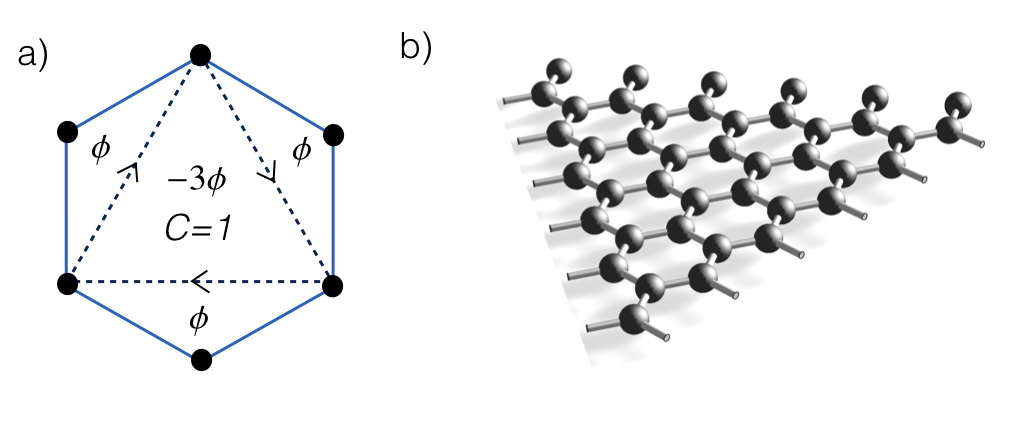}{fig3}{(a) Haldane's tight-binding model after \citep{haldane88}
on a honeycomb lattice, used to show that topological Chern insulators
can form without a net magnetic field. (b) Graphene, which together
with Haldane's model, provided stimulus for the discovery of
topological insulators. }

Topological structures in physics can develop in both 
real space and in momentum space.  An example of the first kind of topology, 
are vortices in a superfluid. 
In a superfluid, the phase $\phi (x)$ of the complex order
parameter $\psi (\vec{ x})\propto
e^{i\phi (x)}$ is a smooth function of position and
in passing around a closed path the order parameter must
change smoothly and come back to itself, 
so that the change in the phase must be an integer
multiple $n$ of $2\pi$, $n\times 2\pi$.  The integer $n$
describes the quantization of circulation in a superfluid, first
predicted by Onsager and Feynman. 

A second-kind of topology involves the wavefunction of electrons, in which
a non-trivial topological configuration constitutes
a new kind of ``topological order''
\[
\hbox{differential geometry of the wavefunction }\leftrightarrow
\hbox{topological order}
\]
Topological order is  distinct from 
broken symmetry and it manifests itself
through the formation of gapless surface or edge (2D) excitations
around the exterior of an otherwise insulating state.
The first example of such
topological order is the quantization of the Hall constant in two
dimensional electron gases, according to the relationship
\[
\rho_{xy}= \frac{1}{\nu}\frac{h}{e^{2}}
\]
where the Hall resistivity, 
$\rho_{xy}=V_{H}/I$ is the ratio of the transverse Hall voltage $V_{H}$
to the current $I$ 
and $\nu$, an integer associated with the
the topology of the filled electron bands\citep{tknn}; one of
the manifestations of this effect, is the formation of $\nu$ ``edge states''
which propagate ballistically around the quantum Hall insulator. 

Microscopically, this
topology is determined by way the phase of the electron wavefunction
twists through momentum space, which is given by 
a quantity called the ``Berry connection'' associated with the
filled electron bands, given by 
\[
\vec{\cal A} (\bk) =- i \sum_{m=1,N}\langle u_{m,\bk} \vert  
\nabla_{\bk }\vert u_{m,\bk }\rangle, 
\]
where $u_{m,\bk }$ is the Bloch wavefunction of the mth filled
electron band at momentum $\bk $.   The Berry connection $\vec{\cal A} (\bk
)$ plays the role of an
emergent vector potential: a momentum-space analog of the
electromagnetic field.
The corresponding magnetic flux, or  ``Berry curvature'' 
$\kappa_{\bk }= \vec{\nabla} \times \vec{\cal A} (\bk)$ plays the same role as the curvature in the Gauss-Bonnet theorem,
and the integral of this curvature over momentum space gives the
integer 
``Chern number'', 
\begin{equation}\label{}
\nu = \frac{1}{2\pi}\int 
\kappa_{\bk}d^{2 }k.
\end{equation}

Later in the 1980s, Duncan Haldane 
showed that such topological order
could occur without a net magnetic field\citep{haldane88}. Haldane's
1987 theoretical model had a honeycomb structure (Fig. \ref{fig3}a.). 
Fifteen years later, 
the discovery of a 2D
carbon structure ``graphene'', with an uncanny resemblance to Haldane's 
model, inspired Charlie Kane
and Eugene Mele\citep{kanemele05} to propose that 
topological order would develop in graphene without any magnetic
field (Fig. \ref{fig3}b.). The key to their idea was ``spin-orbit''
coupling - an internal magnetic coupling
between the spin and orbital motion of electrons. 
Kane and Mele recognized that spin-orbit coupling allows spin-up
electrons to create a magnetic field for spin-down electrons, and vice
versa, creating two separate versions of the quantum Hall effect,
one for spin- up and one for spin-down  electrons. The resulting edge states 
carry spin, forming an early version of 
the modern topological insulator, the ``spin-Hall insulator''. 

Although the spin-orbit coupling in real graphene turned out to be 
too 
weak to give rise to a topological insulator, the idea held and was
confirmed by experiment\citep{bernevig06,konig07} and
later generalized to the three dimensional topological
insulators\citep{kane1,kane2,rahulroy09}.  The current view is 
that spin-orbit coupling changes the topology of an insulator
by inducing a crossing between the unoccupied conduction and occupied
valence bands.  Such crossings can only take place at certain allowed 
high symmetry points in momentum space
defined by the crystal symmetry, and when they do, they change the
topology. Like the braiding of a
ribbon, where an odd number of twists produces a non-trivial
configuration or M\"obius strip, in insulators, 
an odd number of band crossings leads to a ``strong
topological insulator'' (STI) with conducting surface states. 

From a fully reductionist viewpoint, one might wonder 
wonder why the topological 
revolution did not occur along with the development of
electron band theory, from which it can be deduced. Indeed, one of the
early pioneers of band theory, the co-inventor of the transistor,
William Shockley\citep{Shockley:1939zz}, came remarkably close.
Yet new emergent principles, while traceable back
to their microscopic origins, required the experimentally-inspired
development of new concepts.  We
see here a
close analogy with the 200 year delay in the discovery of energy as a
consequence of Newtonian mechanics.

\subsection{Strange Metals}\label{}

As a counterpoint to the discovery of topological insulators, 
I'd like to say a little about how our
understanding of metals appears to be on the verge of radical change. 
The foundations of the modern theory of metals were 
established 
not long after 
the discovery of the electron, at the turn of the 20th century
by Paul Drude.  
One of the main ideas of Drude's theory of metals, is that
electrons defuse through a metal, due to their scattering off
imperfections and vibrations.  The resulting ``transport relaxation time''
$\tau_{tr}$ governs most aspects of the electron transport.
The arrival of quantum mechanics in the 1920s led to a major upheaval
in the understanding of the electron fluid. In particular, electrons,
as identical quantum particles, were found to obey the Pauli Exclusion principle, which
prevents more than one of them occupying the same eigenstate. This
individualism causes electrons to fill up momentum space to higher
and higher momentum states up to some maximum momentum, the Fermi
momentum. The occupied states at this maximum momentum define a “Fermi
surface” in momentum space, and almost  all the action in a metal involves electrons at the Fermi surface.
\figwidth=\textwidth
\fg{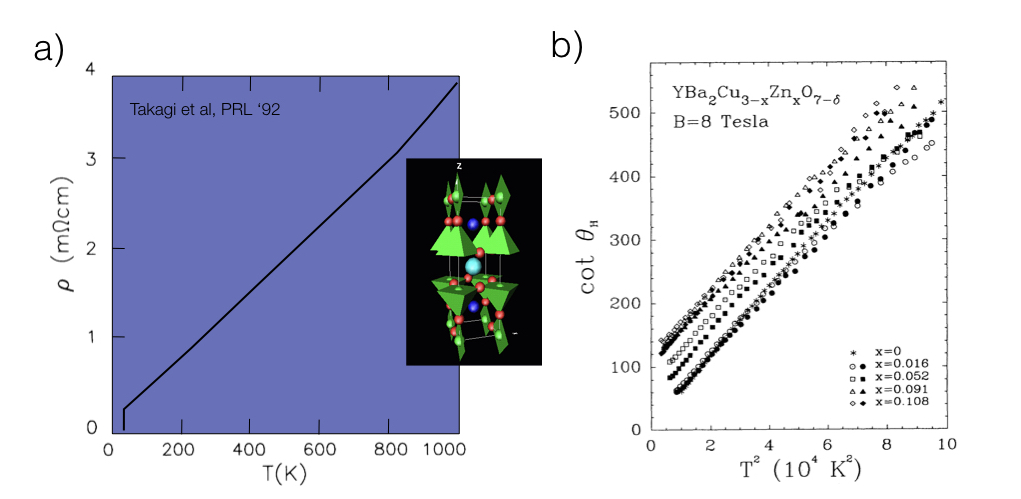}{fig4}{Strange metals.  (a) Linear resistivity of the
high temperature superconductor La$_{2-x}$Sr$_{x}$CuO$_{4}$ (x=0.15)
(adapted with permission from H. Takagi {\sl et al}, 
Phys. Rev. Lett. 69, 2975 (1992) 
\citep{takagi92}), showing
the remarkable linear resistivity up to 1000K, indicating 
that the electrical current relaxation rate
$\Gamma_{tr}\propto T$ is proportional
to the temperature. (b)  Quadratic temperature dependence of the Hall
angle in a cuprate superconductor (reprinted with permission from  T. R. Chien, {\sl et
al.}, Phys. Rev. Lett. 67, 2088 (1991)\citep{ong92}), indicating that Hall currents in these strange metals
exhibit a decay rate $\Gamma_{H}\propto T^{2}$. 
The appearance of two relaxation time-scales in a simple
conductor poses a challenge to our current understanding of metals.}

Yet when the dust of quantum mechanics settled, 
Drude's picture had survived
almost intact: in particular, the concept of a transport relaxation time
could be extended to describe the scattering of electrons at the Fermi
surface by disorder and mutual interactions, leading back to Drude's  diffusive
electron transport picture. One of the consequences of this
robustness, is that one can measure the resistivity, Hall constant and
the dependence of its resistivity on a magnetic field, the so-called
``magneto-resistance'', to check if these quantities scale
with the scattering time $\tau_{tr}$ in  the way predicted by
Drude theory.  Although the rate at which electrons scatter is
temperature dependent, various ratios appearing in the transport
theory are independent of the scattering rate and become temperature
independent. 
One well-known  consequence of Drude theory, is that the
cancellation between the scattering rate associated with the Lorentz
force cancels with the scattering rate due to the electric force, so
that the ratio of the two, determined by the Hall  constant $R_{H}=
V_{H}/I$ is temperature independent. Another consequence
is a scaling law known as Kohler's law. In Drude theory the resistivity
$R$ is proportional to the scattering rate $R\propto
\tau_{tr}^{-1}$, whereas the magneto-resistivity grows with 
the square of the 
angle of deflection (Hall angle) of the electrical current in a field,
$\Delta R/R\propto \Theta_{H}^{2}$.  Now the Hall angle depends on the
product of the cyclotron precession frequency and the scattering rate,
$\theta_{H}= \omega_{c}\tau_{tr}$, so that 
 $\Delta
R/R \propto \theta_{H}^{2}\sim \tau_{tr}^{2}$, which when combined
with the resistivity, leads to Kohler's rule
$\Delta R/R \propto R^{-2} $.  This scaling
relation works remarkably well for a wide range of simple metals,
vindicating Drude's theory.

Of course, Quantum mechanics does have radical consequences for
metals. For example, disorder can cause electron waves to {\sl Anderson
localize}\citep{anderson58,gangoffour}, completely 
stopping electron diffusion to produce an insulator. The many-body version of this phenomenon, {\sl many-body
localization}\citep{manybodylocalization} is of great current interest.
Another radical consequence that I want to discuss now, 
is the formation of {\sl strange metals}.
Over the past three decades, experiments  have revealed a
new class of ``strange metal'' which deviates from Drude theory in a
qualitative way.   This unusual metallic behavior tends to develop in
metals that are close to instability. When the interactions are
increased inside a metal, through the effect of pressure, chemistry or
external fields, the metal can become unstable, 
giving rise to {\sl Quantum Phase transition} into an ordered state,
such as magnetism.  Such instabilities, 
occur at a absolute zero, where there are no
thermal fluctuations to drive a phase transition. Instead, the phase
transition is driven by quantum zero point fluctuations, and it is
thought that these
fluctuations play a role in transforming the electron fluid 
causing the resulting conductor 
to deviate qualitatively from Drude behavior. 

The most famous  strange metals
are the high temperature cuprate superconductors\citep{takagi92,ong92}, but similar
behavior is also seen in their low temperature cousins, the family of
heavy electron superconductors known as ``115'' superconductors\citep{Nakajima:2004wy}, and
most recently, in artificially constructed two dimensional electron
gases\citep{Mikheev:2015cb}, which are not superconducting. 
High temperature cuprate superconductors 
lose their resistance at temperatures as high as 90K, high 
enough to be able to use liquid nitrogen to cool them into the
superconducting state. 
But above these temperatures, they are equally
remarkable, for they exhibit a resistivity that is linear up to very
high temperatures $R (T)\propto T$(Fig. \ref{fig4}a.). In fact, this linearity can be
traced back to a Drude scattering rate $\Gamma_{tr}=\tau_{tr}^{-1}$ that is proportional to the
temperature, given approximately by  $\Gamma_{tr}\sim \frac
{k_{B}T}{\hbar }$.
The  time-scale $\tau_{tr}\sim \frac{\hbar }{k_{B}T}$ is sometimes called the ``Planck time'', because it is the
time scale derived from combining the energy-time uncertainty relationship
$\Delta E \Delta \tau \sim \hbar $ with the Boltzmann 
energy $\Delta E \sim k_{B}T$. This simple scaling 
of the scattering rate with the temperature is very unusual, and in a
typical metal the scattering rate
has a much more complicated dependence on temperature, on disorder and
on the coupling to vibrations of the crystal.  
Perhaps the strangest aspect of these metals, is  
their departure from Drude behavior in a magnetic field
because the scattering response to the Lorentz force, 
measured in a magnetic field is
qualitatively different to the response to a pure electric field. 
Whereas the linear resistivity
gives a scattering rate $\Gamma_{tr}$ proportional to temperature, the
magneto-resistivity and Hall resistivity give a scattering rate that
is quadratic in the field $\Gamma_{H}\propto \theta_{H}^{-1}\propto
T^{2}$ (Fig. \ref{fig4}b).  Summarizing
\[
\Gamma_{tr}\propto T, \qquad \Gamma_{H}\propto \frac{T^{2}}{W},
\]
where $W$ is a scale that governs the decay of Hall currents. 
The presence of these qualitatively different scattering rates leads
to a strongly temperature dependent Hall constant, and a 
``modified Kohler's rule'', whereby the
magneto-transport scales with the square of the Hall angle, rather
than the square of the conductivity, 
\begin{equation*}
R (T)\propto \tau_{tr}^{-1}, \qquad \frac{\Delta R}{R} \propto \theta_{H} (T)^{2}
\end{equation*}
This behavior is not unique to cuprate
superconductors, and it has also been observed in certain heavy
fermion superconductors\citep{Nakajima:2004wy}, which are low temperature cousins of the
cuprate superconductors, and in low dimensional oxide interfaces\citep{Mikheev:2015cb},
but which have quite different microscopic
chemistry and structure. 
These results taken together suggest that a
fundamentally new kind of metal has been discovered, one that may
require a new conceptual framework for interacting electrons.
Unlike the new developments in our understanding of insulators,
the emergent framework for understanding  strange metals is still very
much in the early days of discovery. 

\section{Conclusion}\label{}

This article has illustrated examples of 
of emergence in condensed matter physics, 
seeking to highlight the close interdependence of a reductionist and emergent
approach. 
Perhaps the most exciting aspect of this linkage, is that it 
may provide a 
way to accelerate the way we solve major problems in
physics and the Natural sciences.  While reductionism provides the
mathematics and the computational tools to tackle complex problems and
gain new insight into emergence, at the same time, it is likely that
the importane of understanding of physics in the lab, particularly 
emergent physics between the Angstrom and the Micron, will, as it has
in previous centuries,  yield important insights into our reductionist
understanding. 

As in
previous generations, condensed matter physicists are looking to the
tools of particle physics, such as the holographic principle\citep{zaanen}, to make 
new progress on the many body problem, while in a similar
vein, particle physicists and cosmologists are  looking to emergence
and condensed matter for inspiration.   One of the prevalent ideas for
unifying gravity and quantum mechanics is that space-time itself may be
an emergent property of quantum gravity on scales beyond the Planck
length\citep{seiberg}. Another area of activity is the problem of dark
matter. For example, recently Verlinde\citep{Verlinde:2016tj}
has suggested that the dark matter problem may be a consequence of an
emergent aspect of gravity in which the unseen gravitating force
inside galaxies is not interpreted as a cloud of particles, but
as a kind of gravitating condensate.

These developments  tempt us to speculate whether our current
understanding of quantum mechanics
might parallel that of classical
mechanics, which remained incomplete 200 years after
Principia, because one of its key emergent consequences, heat, prevented an
understanding of energy. Perhaps, in a similar fashion, 
90 years after Heisenberg, Schr\"odinger and
Dirac, a more complete understanding of
quantum mechanics might await new perspectives on emergence. 

%% Definition of emergence?
%% I have stayed away. 
%% Tried to emphasize: Link between the cosmos and the lab.
%% Reductionism and emergence.

%\section{Introduction}\label{} 

%
%Quick attempt at an outline.
%

I should like to thank 
Natan Andrei, 
Stephen Blundell, Tom Lancaster  and Jan Zaanen for stimulating discussions related to this work.
This work was supported by NSF grant DMR-1309929. 

\bibliographystyle{agsm}
\bibliographystyle{rmp}
%\bibliography{durhamrefs}
%merlin.mbs apsrmp4-1.bst 2010-07-25 4.21a (PWD, AO, DPC) hacked
%Control: author (3) reversed first dotless

%

\end{document}